\documentclass{article}
\usepackage{spconf}
\usepackage{amsmath}
\usepackage{amssymb}
\usepackage{graphicx}
\usepackage{caption}
\usepackage{subcaption}
\usepackage{comment}

\def\x{{\mathbf x}}

\title{Target Tracking with Electrical Impedance Tomography}
\name{Timo Huuhtanen$^{\star}$ \qquad Antti Lankinen$^{\dagger}$ \qquad Alexander Jung$^{\star}$}

\address{$^{\star}$ Department of Computer Science, Aalto University, Finland\\
	$^{\dagger}$Department of Bioengineering, Imperial College, United Kingdom%\\
}
\begin{document}
%\ninept
%
\maketitle
\begin{abstract}
Electrical impedance tomography (EIT) has been successfully applied to several important application domains such as medicine, geophysics and industrial imaging. EIT offers a high temporal resolution, which allows to track the location of a moving target on a conductive surface  accurately. Existing EIT methods are geared towards high image quality instead of smooth target trajectories, which makes them suboptimal for target tracking. We combine EIT methods with hidden Markov models for tracking moving targets on a conductive surface. Numerical experiments indicate that the proposed method outperforms existing EIT methods in target tracking accuracy.
\end{abstract}
\section{Introduction} \label{sec:introduction}

Electrical impedance tomography (EIT) is a non-intrusive method for 2- or 3-dimensional imaging of a conductive body. Electrodes that are connected to the body are used to inject electrical currents and measure the resulting voltages. This procedure is repeated for a set of specified current patterns. The resulting set of voltage measurements is used to create an image of the conductivity distribution of the body by solving the corresponding inverse problem. 

EIT has been used in medical, industrial and geophysical applications for bodies that are static or only slowly varying.  EIT's high temporal resolution makes it potential for sensing surface applications \cite{zhang_electrick:_2017}, \cite{yoshimoto_tomographic_2020}. In this paper, we apply EIT to tracking the location of a moving target on a 2-dimensional conductive surface. The requirements of target tracking differ from those of static imaging applications that aim at preservation of details, edges and shapes in the image. When tracking a target on a surface, the location of the target and accurate tracking of its movement are important. Therefore, traditional EIT algorithms are suboptimal for this application as indicated by our simulation results in Section \ref{sec:experiments}.

Detection of the location of a single small-sized target in EIT has been studied in \cite{lee_non-iterative_2015}. However, there the targets are static and temporal information is not utilized. Temporal EIT methods have been compared in \cite{adler_temporal_2007}. Reconstructing the image for each time instance independently tends to result non-smooth trajectories for moving targets. Improved temporal EIT image reconstruction methods have been proposed: \cite{vauhkonen_kalman_1998} proposes the use of Kalman filters, \cite{kim_image_2001} uses extended Kalman filters, and \cite{adler_temporal_2007} uses the measurement data from several time instances to reconstruct the image for one instance of time. All these methods operate with EIT voltage measurements, where utilization of detailed target movement information is difficult.

Our approach combines the EIT inverse problem with a model of the target movement (trajectory). The trajectory of a moving target is smooth and the speed is limited (the precise limit depending on the application). The proposed method models the moving target with hidden Markov model (HMM) and finds the most likely trajectory with Viterbi algorithm.  Kalman filter methods (\cite{kim_image_2001}, \cite{vauhkonen_kalman_1998}) are also based on HMMs, but there HMM is used to model the conductivity, and the location of the target is found with postprocessing. In our method, the state of HMM is an index of the mesh directly indicating the location of the target. We compare the performance of our algorithm with state-of-the-art EIT temporal image reconstruction algorithms on synthetic data. Our method outperforms traditional EIT reconstruction methods in target tracking accuracy.

The remaining part of this paper is organized as follows. In Section \ref{sec:EIT_setup}, we describe EIT measurement setup. In Section \ref{sec:methods}, we formulate the temporal EIT image reconstruction problem and the compared methods. The numerical experiments are discussed in Section \ref{sec:experiments}. Section \ref{sec:conclusion} presents a conclusion and discusses follow-up research directions.

\textbf{Notation:} Boldface lowercase (uppercase) letters denote vectors (matrices). The transpose of a matrix $\mathbf{X}$ is denoted by $\mathbf{X}^{T}$. Gradient is denoted by $\nabla$. We denote the probability of event $x$ by $P(x)$ and the probability density function of random variable $X$ by $p_{X}$. For a vector $\mathbf{x} \in \mathbb{R}^{n}$, $\mathbf{x}_{+} = [\max(x_{1}, 0) \hdots \max(x_{n}, 0)]^{T}$. We also need the diagonal of a matrix:
\begin{equation*}
 \mathrm{diag}: \mathbb{R}^{n \times n} \rightarrow \mathbb{R}^{n \times n}, \mathrm{diag}(\mathbf{X}) =
 \begin{bmatrix}
  \mathbf{X}_{1,1} &   &  \\
   & \ddots &  \\
   & & \mathbf{X}_{n,n}
  
 \end{bmatrix}.
\end{equation*}
\section{EIT setup} \label{sec:EIT_setup}

\begin{figure}[htbp]
	\includegraphics[width=0.46\textwidth]{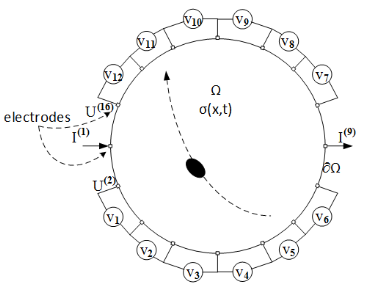}
	\caption{EIT principle: electrical current patterns are applied to electrodes attached to the boundary $\partial\Omega$ and voltages are measured using the same electrodes. Voltages are obtained as potential differences between two electrodes. Currents and voltages allow to deduce changes to the conductivity, $\sigma$, caused by the target moving on the conducting surface $\Omega$.}
	\label{fig:EIT_meas}
\end{figure}

A conductive surface $\Omega$ with inhomogeneous and time-varying conductivity $\sigma$  is probed with L electrodes at positions $e_{1} \hdots e_L$. Electrical currents, $I^{(l)}$, are imposed to the electrodes and the resulting voltages, $U^{(l)}$, are measured using the same electrodes as illustrated in Fig. \ref{fig:EIT_meas}. Potential distribution in $\Omega$ is denoted as $u$. Voltages and currents are coupled as \cite{mueller_linear_2012},
 
\begin{equation}
\label{eq:divergence}
\nabla \cdot (\sigma \nabla u) = 0, x \in \Omega
\end{equation}

\begin{equation}
\label{eq:U_border}
u + z_l \sigma \frac{\partial u}{\partial \nu} = U^{(l)}, x \in e_l, l=1,2,...,L
\end{equation}

\begin{equation}
\label{eq:I_border}
\int_{e_l} \sigma \frac{\partial u}{\partial \nu} dS = I^{(l)}, x \in e_l, l=1,2,...,L
\end{equation}

\begin{comment}
\begin{equation}
\label{eq:U0_border}
\sigma \frac{\partial u}{\partial \nu} = 0, x \in \partial \Omega \setminus \cup_{l=1}^{L} e_{l}
\end{equation}
\end{comment}

Here, (\ref{eq:divergence}) indicates that there are no current sources or sinks inside of $\Omega$. (\ref{eq:U_border}) tells that the measured voltage on the electrode is a sum of two terms: voltage on the boundary of the surface and voltage due to the contact impedance between $l^{\mathrm{th}}$ electrode and the surface, $z_l$; here $\nu$ is outward unit normal on the boundary. (\ref{eq:I_border}) defines the dependency between the currents on the electrodes and the current density on the boundary of the surface.

\begin{comment}
, and (\ref{eq:U0_border}) states that there is no current flow across the surface boundary except via the electrodes. 

Without loss of generality, we can make the following assumptions for the electrode voltages and currents
\begin{equation}
\label{eq:U_sum}
\sum_{l=1}^{L} U^{(l)} = 0
\end{equation}

\begin{equation}
\label{eq:I_sum}
\sum_{l=1}^{L} I^{(l)} = 0.
\end{equation}

If $\sigma$ is known, and the currents $I^{(l)}$ are applied to the surface via the electrodes, $u$ can be calculated based on (\ref{eq:divergence})-(\ref{eq:U0_border}). This is called forward problem of EIT. If 4-electrode model is used in the measurement, it means that two electrodes are used for current source and sink and the rest are used for voltage measurements. Due to (\ref{eq:U_sum}) there are $L-3$ independent voltage measurements on the border, $\partial\Omega$, for each of the $L$ current patterns. The measured voltages, $U^{(l)}$ (\ref{eq:U_border}) can be collected to a $L(L-3)$-by-$1$ vector, $\mathbf{v}$.
\end{comment}

We consider an EIT system with $L$ electrodes on the boundary of a 2D unit circle. Using the electrodes, $L$ drive current patterns are applied. Voltage differences between adjacent electrodes are measured excluding the electrodes where currents are applied, and $n_{V} = L - 4$. The voltage measurements from $L$ patterns are collected into a vector $\mathbf{v} \in \mathbb{R}^{n_M}$, where $n_M = n_V L$. We consider difference EIT where voltages and conductivities are treated as differences against those measured with an empty sensing surface. Data points are then $\mathbf{y} = \mathbf{v} - \mathbf{v}_{0}$, where the reference voltage measurements, $\mathbf{v}_{0}$, are measured from an empty sensing surface with conductivity $\boldsymbol{\sigma_0}$. We assume that $\mathbf{v}_{0}$ is noise free as it can be averaged from multiple measurements when the surface is known to be empty. We denote the difference data collected at time $t$ with $\mathbf{y}(t)$.

\begin{figure}[htbp]
	\includegraphics[width=0.46 \textwidth]{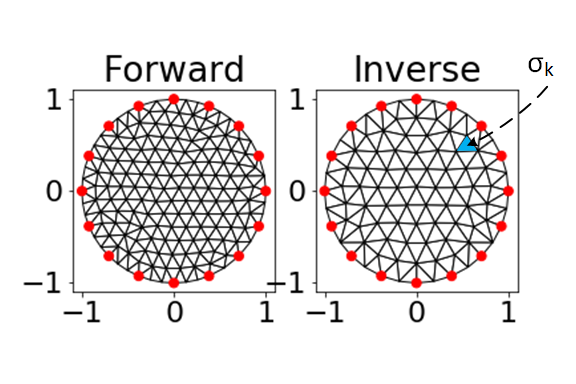}
	\caption{Finite element meshes used in forward and inverse problems.}
	\label{fig:distmesh}
\end{figure}

The EIT inverse problem is

\begin{equation}
\underset{\boldsymbol{\sigma} }{\arg\min} \|\mathbf{v} - \mathbf{u}(e_1, \hdots, e_{L},\boldsymbol{\sigma}) \|_{2}^{2}
\label{eq:nonlinear_inv}
\end{equation}

where $\mathbf{u}(e_1, \hdots, e_{L},\boldsymbol{\sigma}) \in \mathbb{R}^{n_M}$ is the concatenation of solutions to the governing partial differential equation (PDE) at the electrode locations for each current pattern. The problem is typically made tractable by modeling the surface as a finite element mesh with $n_N$ triangular elements. Conductivity distribution is expected to be constant within each element; these values are the elements of vector $\boldsymbol{\sigma} \in \mathbb{R}^{n_N}$. Fig. \ref{fig:distmesh} illustrates possible meshes for the finite element method (FEM). The PDE can then be linearized at an estimated reference conductivity $\boldsymbol{\sigma}_{0}$:

\begin{equation}
 \mathbf{u}(e_1, \hdots, e_{n_V}, \boldsymbol{\sigma}) = \mathbf{u_0} + \mathbf{J}(\boldsymbol{\sigma} - \boldsymbol{\sigma_0}).
 \label{eq:linearisation}
\end{equation}

Here $\mathbf{J}$ is the Jacobian where $J_{ij}=\frac{\partial u_i}{\partial \sigma_j}$.

We denote the differential conductivity $\boldsymbol{\sigma} - \boldsymbol{\sigma}_0$ at time $t$ by $\mathbf{x}(t)$, where each element $x_i(t)$ corresponds to conductivity difference value at one triangular element of the inverse mesh. Inserting (\ref{eq:linearisation}) into (\ref{eq:nonlinear_inv}),

\begin{equation}
 \underset{\mathbf{x}}{\arg\min} \|\mathbf{y} - \mathbf{Jx} \|_{2}^{2} .
\end{equation}

This system is typically underdetermined $(n_N > n_M)$ and regularization is required. A standard choice is $L_2$ regularization:

\begin{equation}
 \underset{\mathbf{x}}{\arg\min} \|\mathbf{y} - \mathbf{Jx} \|_{2}^{2}  + \lambda \| \mathbf{Rx} \|_{2}^{2}
\end{equation}

where $\lambda$ is a hyperparameter and $\mathbf{R}$ is a regularization matrix containing some prior information about $\mathbf{x}$. If all the elements of $\mathbf{x}$ are assumed to be independent and have equal expected value, $\mathbf{R}$ is the identity matrix $\mathbf{I}$. In EIT, such solutions tend to push reconstructed noise to the boundary as the measured data are much more sensitive to boundary elements \cite{adler_temporal_2007}. $\mathbf{R}$ can be scaled with the sensitivity of each element, so that $\mathbf{R} = (\mathrm{diag}(\mathbf{J}^{T} \mathbf{J}))^{p}$ for some exponent $p \in [0, 1]$. For $p=1$, this is the NOSER prior \cite{cheney_noser:_1990}. As in \cite{adler_temporal_2007}, in this paper we use $p=0.5$ as a heuristic compromise between pushing noise to the boundary $p=0$ and center $p=1$.

The problem now reads
\begin{equation}
  \mathbf{\hat x} = \underset{\mathbf{x}}{\arg\min} \|\mathbf{y} - \mathbf{Jx} \|_{2}^{2}  + \lambda \| \mathrm{diag}(\mathbf{J}^{T} \mathbf{J}))^{0.5} \mathbf{x} \|_{2}^{2}
\end{equation}

and the solution can be obtained as
\begin{align}
 \mathbf{\hat x} &= (\mathbf{J}^{T} \mathbf{J} + \lambda\ \mathrm{diag}(\mathbf{J}^{T} \mathbf{J})^{0.5})^{-1} \mathbf{J}^{T} \mathbf{y} \\
 &= \mathbf{Hy}.
 \label{eq:gn_solution}
\end{align}

\section{Methods} \label{sec:methods}

In this paper we compare three target tracking methods: 
\begin{itemize}
	\item Nontemporal method solving EIT image reconstruction problem independently for each time using (\ref{eq:gn_solution})
	\item The method based on Kalman filter \cite{vauhkonen_kalman_1998} 
	\item Our novel method based on HMM of the moving target
\end{itemize}
For simplicity, we only consider targets that have a higher conductivity than the surface itself.

\subsection{Kalman filter}
Applying (\ref{eq:linearisation}) to Kalman filter \cite{kalman_new_1960}, we get 

\begin{align}
 \mathbf{x}(t+1) &= \mathbf{Ax}(t) + \mathbf{n}_{x}(t) \\ 
 \mathbf{y}(t) &= \mathbf{Jx}(t) + \mathbf{n}_{y}(t)
\end{align}

where $\mathbf{A} \in \mathbb{R}^{n \times n}$ is the state transition matrix. $\mathbf{n}_{x}(t) \sim \mathcal{N} (\mathbf{0}, \boldsymbol{\Sigma}_{\mathbf{x}})$ and $\mathbf{n}_{y}(t) \sim \mathcal{N} (\mathbf{0}, \boldsymbol{\Sigma}_{\mathbf{y}})$ are noise in the state transition and observation models, respectively. We can assume that $\boldsymbol{\Sigma}_{\mathbf{x}} = \mathbf{I}$ without loss of generality \cite{roweis_unifying_1999}. In addition, we can assume that the measurement noise is uncorrelated, so $\boldsymbol{\Sigma}_{\mathbf{y}}$is diagonal. For simplicity, we assume $\boldsymbol{\Sigma}_{\mathbf{y}} = \mathbf{I}$.

The choice of $\mathbf{A}$ is typically the identity matrix. However, this choice may not be suited for tracking moving targets as it assumes that the targets are stationary in time. To make the model more suitable for tracking movement, let $\mathbf{S} \in \mathbb{R}^{n \times n}$ be the adjacency matrix of the mesh elements such that $\mathbf{S}_{i,j} = 1$ if elements $i, j$ share an edge or a node and 0 otherwise. Let $\mathbf{D} \in \mathbb{R}^{n \times n}$ be the diagonal degree matrix such that $\mathbf{D}_{i,i} = \Sigma_{k=1}^{n} \mathbf{S}_{k,i}$. Then the transition matrix is given by

\begin{equation}
 \mathbf{P} = (\mathbf{D} + \mathbf{I})^{-1} (\mathbf{S} + \mathbf{I}).
 \label{eq:transition_matrix}
\end{equation}

Informally, this transition matrix assumes that the state spreads out over time, since the probabilities of the target staying in the same location and moving to one of its neighbors are the same. We assume that the movement of the target is slow in comparison to the EIT sampling rate, justifying the use of this transition matrix.

\subsection{Hidden Markov model for EIT}
Let $M$ denote the number of elements in the inverse solution mesh. The HMM consists of two random variables $Q(t) \in [M]$ (state), $Z(t) \in \mathbb{R}^{L^2 - 4L}$ (observation). In our context, $Q(t)$ corresponds to the location of the target at time $t$ and $Z(t)$ is the voltage measurement at time $t$, $\mathbf{y}(t)$. We can solve the most likely sequence of states with the Viterbi algorithm \cite{viterbi_error_1967}. For $T$ measurements, the parameters of the HMM required for the Viterbi algorithm consist of a transition matrix $\mathbf{P} \in \mathbb{R}^{M \times M}$ and emission matrix $\mathbf{E} \in \mathbb{R}^{M \times T}$ with $\mathbf{P}_{i,j} = P(Q(t+1) = i | Q(t) = j)$ and $\mathbf{E}_{i,j} = P(Z(j) | Q(j) = i)$.

The transition matrix is same as used for the Kalman filter and given by (\ref{eq:transition_matrix}). For the emission probabilities, we need to compute $P(\mathbf{z} | q)$ for all states $q$ and voltage measurements $\mathbf{z}$. Using Bayes' theorem:
\begin{equation}
    P(\mathbf{z} | q) = \frac{P(q | \mathbf{z} ) P(\mathbf{z} )}{P(q)}.
\end{equation}

Here, $P(q)$ is the probability of state $q$ in the stationary distribution of $\mathbf{P}$. For $P(q | \mathbf{z})$, we argue that the solution to the inverse problem can be interpreted as a probability distribution for the position of the target knowing that the target spans a single element in the mesh. To complete the interpretation, the solution needs to be normalized:

\begin{equation}
 \mathbf{\tilde \x} = \mathbf{\hat \x}_{+} / \|\mathbf{\hat \x}_{+}\|_{1}
 \label{eq:normalisation}
\end{equation}

where $\mathbf{\hat x}$ is the solution of equation (\ref{eq:gn_solution}).

The quantity $P(\mathbf{z})$ is the probability of voltage measurement $\mathbf{z}$ and it cannot be computed as the voltage measurements come from a continuous distribution. However, the value of the probability density function (PDF) of the distribution at $\mathbf{z}$ gives a relative measure of probability. We modelled the PDF as a Gaussian mixture model learned from the data. Let $p_{Z}$ denote the estimated PDF of $Z$, $\boldsymbol{\pi}$ the stationary distribution of $\mathbf{P}$, and $\mathbf{\tilde \x}$ as in (\ref{eq:normalisation}). Then

\begin{equation}
 P(\mathbf{z} | q) = \frac{\tilde x_{q} p_{Z}(\mathbf{z})}{\pi_{q}}
\end{equation}

and

\begin{equation}
 \mathbf{E}_{i,j} = \frac{((\mathbf{Hy}(j))_{+})_{i}  p_{Z}(\mathbf{v}_{j})}{\| (\mathbf{Hy}(j))_{+} \|_{1} \pi_{i}}.
\end{equation}

With the above formulations, we can compute the most likely sequence of states with the Viterbi algorithm. The prior state probability distribution used is the stationary distribution if no prior information is available. When applying the algorithm for real-time tracking, the prior state can be obtained from the last state of the previous sequence.

\section{Numerical experiments} \label{sec:experiments}

We compared the performance of the three methods for tracking a single small target. The simulations were conducted using pyEIT \cite{liu_pyeit:_2018}. The forward and inverse meshes consisted of 287 and 152 triangular elements respectively, shown in Fig. \ref{fig:distmesh}. 16 electrodes were placed equispaced on the boundary of the surface. We used opposite current pattern with a total of 16 independent current patterns of 12 voltage measurements each. The target was modelled as spanning a single mesh element and randomly moving to an adjacent element at each time step. The baseline conductivity was set at 1 S/m and the conductivity of the target at 1000 S/m.

The position of the target was initialized randomly and randomly moved to an adjacent element of the previous location at each time step. The forward problem was solved independently for each frame. The number of frames used was 500, as it was a sufficient number for the information about the initial position of the target to mostly disappear $(|\lambda_{2}|^{500} \leq 0.01$, where $\lambda_{2}$ is the eigenvalue of $\mathbf{P}$ with the second largest absolute value). Zero-mean Gaussian noise was added to the forward voltage measurements. Five different signal-to-noise ratios were used: 100 dB, 80 dB, 60 dB, 40 dB, and 20 dB. A reference voltage (result of forward computation with no target) was subtracted from the noisy measurement voltages (note that the actual signal-to-noise ratio is therefore lower) to obtain the difference voltage measurements which were used in the reconstructions. The experiment was repeated 100 times for each noise level.

The performance of each algorithm was evaluated measuring the mean absolute error between the true target center and the predicted target center for each setup, averaged over both time and samples. This metric is also used in \cite{gagnon_comparison_2015}. For the hidden Markov model, the predicted center at time $t$ is the center of the element where the target is at time $t$ in the Viterbi path. For the nontemporal and Kalman filter methods, the target was assumed to be located in the element with the highest predicted conductivity for each frame.

\begin{figure}[h!]
\centering
 \begin{subfigure}[b]{0.23 \textwidth}
  \centering
  \includegraphics[width=\textwidth]{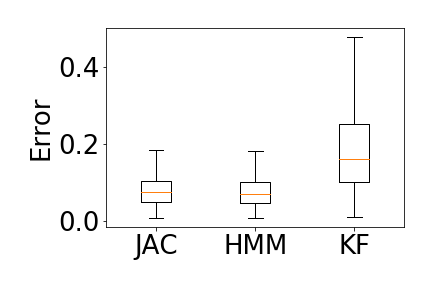}
  \caption{SNR 100 dB}
  \label{fig:snr 100}
 \end{subfigure}
 \begin{subfigure}[b]{0.23 \textwidth}
  \centering
  \includegraphics[width=\textwidth]{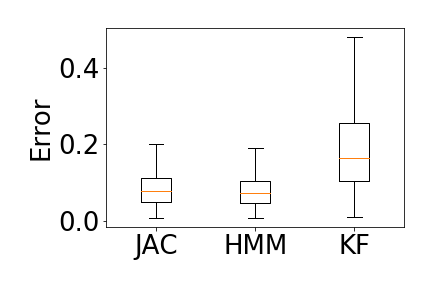}
  \caption{SNR 80 dB}
  \label{fig:snr 80}
 \end{subfigure}
 \begin{subfigure}[b]{0.23 \textwidth}
  \centering
  \includegraphics[width=\textwidth]{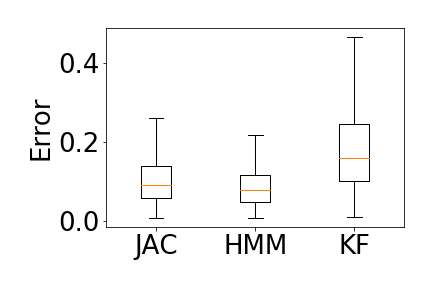}
  \caption{SNR 60 dB}
  \label{fig:snr 60}
 \end{subfigure}
 \begin{subfigure}[b]{0.23 \textwidth}
  \centering
  \includegraphics[width=\textwidth]{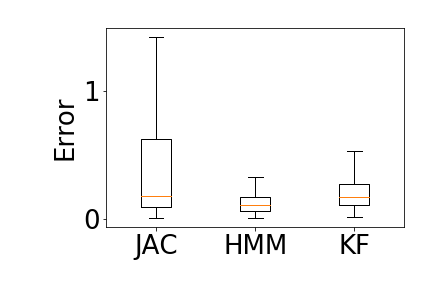}
  \caption{SNR 40 dB}
  \label{fig:snr 40}
 \end{subfigure}
 \begin{subfigure}[b]{0.23 \textwidth}
  \centering
  \includegraphics[width=\textwidth]{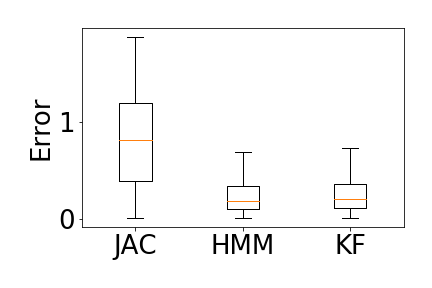}
  \caption{SNR 20 dB}
  \label{fig:snr 20}
 \end{subfigure}
 \caption{Absolute errors of predicted target centers for different algorithms and noise levels. Note the different y-axes on the graphs.}
 \label{fig:experiment}
\end{figure}

\begin{figure}[h!]
 \centering
 \includegraphics[width=0.3 \textwidth]{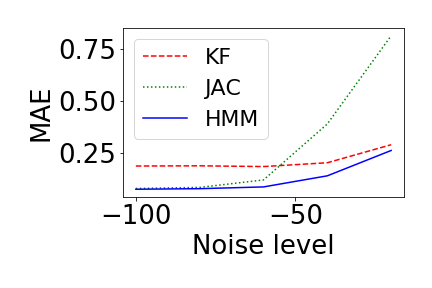}
 \caption{Noise level versus mean absolute error.}
 \label{fig:means}
\end{figure}

The results of the experiment are shown in Fig. \ref{fig:experiment}. It can be seen that our method (HMM) outperforms a simple reconstruction (JAC) and a Kalman filter approach (KF) at all noise levels. In addition, the relative performance of the HMM is consistent across all noise levels. In contrast, simple reconstruction fails at high noise levels and the Kalman filter performs poorly at low noise levels as shown in Fig. \ref{fig:means}.

\section{Conclusion} \label{sec:conclusion}
In this paper, we developed a method for tracking a moving target from electrical impedance tomography data. Our method outperforms existing methods for temporal EIT.

Future research includes tracking of multiple targets, analysis of the lower limit of the EIT sampling rate that preserves the efficiency of the method, and possible extensions to cases where the assumption about a relatively high sampling rate is violated.

\vfill\pagebreak

\bibliographystyle{IEEEbib}
\bibliography{eit}

\end{document}